\documentclass{aastex}
\usepackage{emulateapj5}

\begin{document}
\submitted{}

\title{Slow-roll inflation and CMB anisotropy data}

\author{J\'er\^ome Martin\altaffilmark{1}, Alain Riazuelo\altaffilmark{2}}
\affil{DARC, Observatoire de Paris, UMR 8629 CNRS, 92195 Meudon Cedex, France}
\and
\author{Dominik~J.~Schwarz\altaffilmark{3}} 
\affil{Institut f\"ur Theoretische Physik, Technische Universit\"at Wien, 
       Wiedner Hauptstra\ss e 8--10, 1040 Wien, Austria} 
\altaffiltext{1}{{\tt martin@edelweiss.obspm.fr}} 
\altaffiltext{2}{{\tt Alain.Riazuelo@obspm.fr}}
\altaffiltext{3}{{\tt dschwarz@hep.itp.tuwien.ac.at}}

\begin{abstract} We emphasize that the estimation of cosmological
parameters from cosmic microwave background (CMB) anisotropy data, such 
as the recent high resolution maps from BOOMERanG and MAXIMA-1, requires 
assumptions about the primordial spectra. The latter are predicted from 
inflation. The physically best-motivated scenario is that of slow-roll 
inflation. However, very often, the unphysical power-law inflation scenario 
is (implicitly) assumed in the CMB data analysis. We show that the predicted 
multipole moments differ significantly in both cases. We identify several
misconceptions present in the literature (and in the way inflationary
relations are often combined in popular numerical codes). For example,
we do not believe that, generically, inflation predicts the relation 
$n_{\rm S}-1=n_{\rm T}$ for the spectral indices of scalar and tensor 
perturbations or that gravitational waves are negligible. We calculate the 
CMB multipole moments for various values of the slow-roll parameters and
demonstrate that an important part of the space of parameters $(n_{\rm
S},n_{\rm T})$ has been overlooked in the CMB data analysis so far.
\end{abstract}

\keywords{cosmic microwave background --- early universe} 

\section{Introduction}

Accurate measurements of the cosmic microwave background (CMB)
anisotropies provide an excellent mean to probe the physics of the
early Universe, in particular the hypothesis of inflation. Recently,
scientists working with the BOOMERanG \citep{B98a} and MAXIMA-1 \citep{MAX1a} 
CMB experiments announced the clear
detection of the first acoustic peak at an angular scale $\simeq
1^{\circ }$, which confirms the most important prediction of
inflation: the Universe seems to be spatially flat \citep{B98b,MAX1b}.

In the framework of inflation CMB anisotropies follow from the basic
principles of general relativity and quantum field theory.  To predict
the multipole moments of these CMB anisotropies two ingredients are
necessary: the initial spectra of scalar and tensor perturbations and
the ``transfer functions'', which describe the evolution of the
spectra since the end of inflation. The transfer functions depend on
cosmological parameters such as the Hubble constant ($h$), the total
energy density ($\Omega _0$), the density of baryons ($\Omega _{\rm b}$),
the density of cold dark matter ($\Omega _{\rm cdm}$)
and the cosmological constant ($\Omega_{\Lambda}$).

For the analysis of CMB maps it is a reasonable first step to
test the most simple and physical model of the early Universe:
slow-roll inflation with a single scalar field. Slow-roll inflation
predicts a logarithmic dependence of the power spectra on the wave
number $k$ \citep{S79,M81,G82,S82,H82}. However, in most studies of
the CMB anisotropy the spectral shape of power-law inflation
\citep{AW}, corresponding to an exponential potential for the inflaton
field, has been considered. This case is unphysical, since power-law
inflation does never stop. Two of us \citep{MS2} have shown, using
analytical techniques, that the predictions of power-law and slow-roll
inflation can differ significantly. Here, we confirm these results and
calculate the CMB anisotropies with a full
Boltzmann code developed by one of us (A.~R.). The numerical accuracy
of this code has been tested by comparison to analytical results (low
$\ell$) and to CMBFAST v3.2 \citep{SZ}. In general both codes agree
within $2\%$.
  
We use the new CMB data to test slow-roll inflation, assuming the two
most popular versions of Cold Dark Matter (CDM) models (our
``priors''): the standard CDM model (SCDM: $\Omega_0 =1$, 
$\Omega _{\rm cdm} =1-\Omega _{\rm b}$) 
and the cosmic concordance model ($\Lambda $CDM: $\Omega _0=1$,  
$\Omega_{\rm \Lambda }=0.7$, $\Omega _{\rm cdm} = 1 - \Omega_{\rm b} - 
\Omega_{\rm \Lambda}$), which is motivated by the results of the high-$z$
supernovae searches \citep{P98,R98}. In particular, we take $\Omega_0=1$ 
in agreement with the most important prediction of inflation, $h=0.60$, 
which is consistent with supernovae type Ia measurements [$h = 0.585 
\pm 0.063$ at $90\%$ C.L. \citep{Tammann}], and $\Omega _{\rm b}h^2 = 
0.019$, as inferred from the observed abundance of D and primordial 
nucleosynthesis [$\Omega _{\rm b}h^2 = 0.019 \pm 0.002$ \citep{T00,nucleo}].  

In this letter we recall the basic predictions
of slow-roll inflation (Sec.~2) and correct errors and misconceptions
that have been recently made in the literature on this issue
(Sec.~3). In section 4 we compare for the first time the predictions
of slow-roll inflation with the recent data of BOOMERanG and MAXIMA-1
(without any elaborated statistical technique; we remind that only
$5\%$ of the BOOMERanG data have been analyzed so far).

\section{Predictions of inflation}

The power spectra from power-law inflation, for which the scale factor
behaves as $a(\eta )\propto |\eta |^{1+\beta }$ with $\beta \le -2$,
change with a fixed power of the wavenumber $k$. For the Bardeen
potential and for gravitational waves the power spectra in the
matter-dominated era are respectively given by \citep{AW,MS}
\begin{equation}
\label{specpls}
k^3P_{\rm \Phi}=A \frac{3 - n_{\rm S}}{1 - n_{\rm S}} 
\biggl(\frac{k}{k_0}\biggr)^{n_{\rm S}-1},\  
k^3P_{\rm h}=A \frac{400}{9} 
\biggl(\frac{k}{k_0}\biggr)^{n_{\rm T}},
\end{equation}
where $k_0$ is a pivot scale and where $n_{\rm S}-1\equiv {\rm d}\ln
(k^3P_{\rm \Phi })/{\rm d}\ln k=n_{\rm T}\equiv {\rm d}\ln (k^3P_{\rm h
})/{\rm d}\ln k=2\beta +4$.  The factor $A$ is predicted from
inflation, its expression is given in \citet{MS2}. Here, $A$ is {\it a
priori} free and must be tuned such that the angular spectrum is
COBE-normalized.  The choice of $\beta$ fixes $n_{\rm S}$ and $n_{\rm
T}$ and we always have $n_{\rm S}<1$. The predictions of power-law
inflation are the same for any value of the pivot scale, since $k_0$
can be included into the definition of $A$.

Let us now turn to slow-roll inflation which
is certainly physically more relevant, since it covers a wide class of
inflationary models.  Slow-roll is essentially controlled by two
parameters: $\epsilon \equiv -\dot{H}/H^2$ and $\delta \equiv
-\dot{\epsilon}/(2H\epsilon)+ \epsilon$, where $H$ is the Hubble
rate. These two parameters can be related to the shape of the
inflaton potential \citep{Lid}. All derivatives of $\epsilon$ and
$\delta$ have to be negligible, e.g.~$\dot{\delta}/H = {\cal
O}(\epsilon^2, \epsilon\delta, \delta^2)$, only then the slow-roll
approximation is valid. Slow-roll inflation corresponds to a regime
where $\epsilon$ and $|\delta|$ are constant and small in comparison
with unity. The power spectra of the Bardeen potential and
gravitational waves can be written as \citep{SL,MS2}
\begin{eqnarray}
\label{specsrd}
k^3P_{\rm \Phi} &=& \frac{A}{\epsilon} \biggl[1-2\epsilon -
2(2\epsilon -\delta )\biggl(C+\ln \frac{k}{k_0}\biggr)\biggr], \\
\label{specsrgw}
k^3P_{\rm h} &=& \frac{400A}{9}
\biggl[1-2\epsilon \biggl(C+1+\ln \frac{k}{k_0}\biggr)\biggr],
\end{eqnarray}
where $C\equiv \gamma _{\rm E}+\ln 2-2\simeq -0.7296$, $\gamma _{\rm
E}\simeq 0.5772$ being the Euler constant. Slow-roll inflation
predicts the value of $A$, which is given in \citet{MS2} and has not
necessarily the same numerical value as for power-law inflation.  One
important difference to power-law inflation is that the choice of the
pivot scale $k_0$ now matters. It has been shown in \citet{MS2} that
the slow-roll error in the scalar multipoles is minimized at the
multipole index $\ell =\ell _{\rm opt}$ if $D_{\ell_{\rm opt}}=\ln
(k_0r_{\rm lss})$, where $r_{\rm lss}$ is the comoving distance to the
last scattering surface and $D_{\ell} \equiv 1-\ln 2+\Psi (\ell
)+(\ell +1/2)/[\ell (\ell +1)]$ with $\Psi (x)\equiv {\rm d}\ln \Gamma
(x)/{\rm d}x$. For $\ell _{\rm opt}\gg 1$ this gives $k_0 \simeq
(e\ell _{\rm opt})/(2 r_{\rm lss})$, where $r_{\rm lss} = 2/(a H)_0$
for SCDM and $r_{\rm lss} \simeq 3.3/(aH)_0$ for $\Lambda$CDM. Usually
the choice $k_0 = (aH)_0$ is made, which corresponds to $\ell_{\rm
opt} \simeq 2$. In this letter we also consider the case $\ell _{\rm
opt}=200$, which roughly corresponds to the location of the first
acoustic peak. Finally, from Eqs.~(\ref{specsrd}) -- (\ref{specsrgw})
the spectral indices are inferred
\begin{equation}
\label{specindsr}
n_{\rm S}=1-4\epsilon+2\delta , \quad n_{\rm T}=-2\epsilon .
\end{equation}
An important consequence of these formulas is that the relation
$n_{\rm S}=n_{\rm T}+1$ does not hold for slow-roll inflation, except
in the particular case $\epsilon =\delta$.

\section{CMB data analysis}

\begin{figure}
\begin{center}
\includegraphics[angle=270,width=0.7\linewidth]{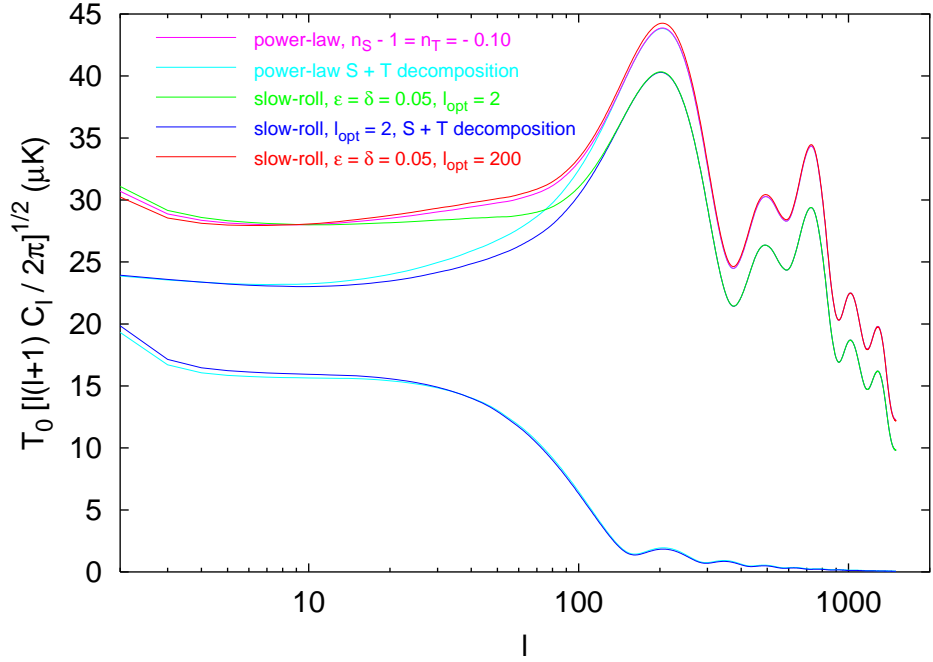}
\end{center}
\caption{
Comparison of CMB band powers from power-law and slow-roll inflation in
the SCDM scenario. The slow-roll model has $\epsilon =\delta =0.050$ such 
that the scalar and tensor spectral indices agree in both cases 
($n_{\rm S}=0.9$, $n_{\rm T}=-0.1$), which means for the power-law model 
$\epsilon = \delta \simeq 0.053$. For $\ell _{\rm opt}=2$, the usual
pivot, the difference between the power-law and slow-roll spectra is large,
which improves for a pivot $\ell _{\rm opt}=200$. The contribution of 
gravitational waves is displayed for power-law and slow-roll inflation 
($\ell _{\rm opt}=2$).}
\label{compsr-pl}
\end{figure}

\begin{figure}
\begin{center}
\includegraphics[angle=270,width=0.7\linewidth]{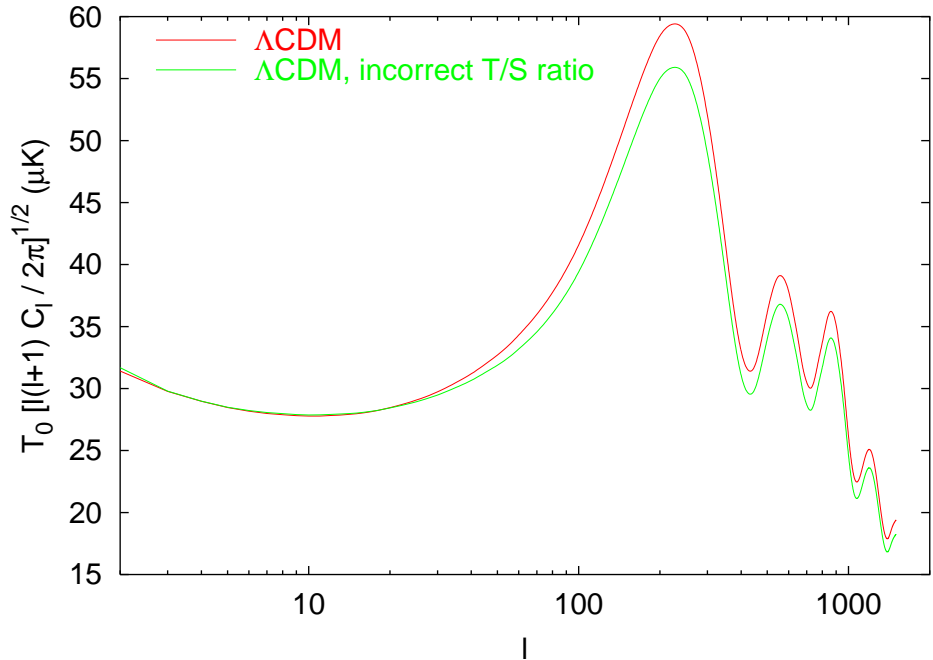}
\end{center}
\caption{
CMB band powers for a power-law spectrum ($n_{\rm S}=0.9$) 
in the $\Lambda$CDM scenario with correct (red line) and incorrect 
(green line) contribution of gravitational waves.}
\label{wrongts}
\end{figure}

We found five misconceptions in the literature, which do have an
important impact on the extraction of cosmological parameters from the
measured CMB multipole moments: a- From Eqs.~(\ref{specpls}) and
(\ref{specsrd}) -- (\ref{specsrgw}) we see that the shapes of the
spectra are not the same in power-law inflation and in slow-roll
inflation (even if $\epsilon =\delta$).  Unfortunately, the unphysical
power-law shape (\ref{specpls}) is assumed frequently, although the
relevance of deviations from the power-law shape has been discussed
earlier [see e.g.~\citet{KT,Lid}]. This difference in the shape
affects the estimates of cosmological parameters in \citet{B98b} and
\citet{MAX1b}, since this misconception is built in into the most
commonly used numerical codes: CMBFAST and CAMB \citep{CAMB}.
In \citet{MS2} it has been demonstrated that the difference is
important and increases with $\vert n_{\rm S}-1\vert$.  
For instance, with the usual choice $\ell _{\rm opt}=2$, the 
error is $15\%$ at $\ell \simeq 200$ for $n_{\rm S}=0.9$, see
Fig.~\ref{compsr-pl}. It has been suggested \citep{MS2} to move the
pivot scale to $\ell_{\rm opt}\simeq 200$, which decreases the difference
from the spectral shapes. For the case considered before, the 
difference reduces to $\simeq 2\%$ with $\ell_{\rm opt}\simeq 200$, as 
can be seen in Fig.~\ref{compsr-pl}. For cases $\epsilon \neq \delta$
the error from the wrong shape increases [for the primordial spectra
this has been studied by \citet{GL1}]. Thus, for the accurate
estimation of the cosmological parameters, one must not mistake
power-law inflation for slow-roll inflation. We suggest to place the
scale for which the slow-roll parameters $\epsilon$ and $\delta$ are
determined in the region of the acoustic peaks, rather than in the
COBE region, which decreases the error from the slow-roll
approximation and one can get rid of the limitations from cosmic
variance for the normalization. 
b- In various publications \citep{B98b,MAX1b} and codes (CMBFAST 
and CAMB) $n_{\rm S} \geq 1$ and $n_{\rm S}-1 \neq n_{\rm T}$ are 
allowed in the data analysis, while working with power-law spectra 
(the prediction of power-law inflation). This is meaningless in the 
context of inflationary perturbations.  For the case $n_{\rm S} = 1$ 
the scalar amplitude is divergent and the linear approximation breaks 
down [see Eq.~(\ref{specpls})]. If, nevertheless, the power--law shape is
assumed, $n_{\rm S} = n_{\rm T} + 1 < 1$ should be fulfilled. On the
contrary, in slow-roll inflation, as can be checked on
Eqs.~(\ref{specindsr}), one can have $n_{\rm S}=1$ or $n_{\rm S}>1$,
only $n_{\rm T} < 0$ is compulsory.  c- A third misconception is that
gravitational waves are not taken into account properly. This is an
important issue since a non-vanishing contribution of gravitational
waves modifies the normalization and changes the height of the first
acoustic peak. In \citet{B98b} (see the footnote [13]), it was assumed
that if $n_{\rm S}>1$, there are no gravitational waves at all, a
supposition in complete contradiction with the predictions of
slow-roll inflation. Also, in that article, the relation $k^3P_{\rm
h}/k^3P_{\rm \Phi}= -(200/9)n_{\rm T}/(1-n_{\rm T}/2)$ was used.  It
is valid for power-law inflation only. In \citet{BZL} gravitational
waves have been neglected, which restricts the analysis for their
choice of $n_{\rm S} = 1$ to the case $\delta = 2\epsilon \ll {\cal
O}(0.01)$, such that tensors contribute less than about $10\%$ of the
power.  d- By default in the CMBFAST and CAMB codes the contribution
of gravitational waves is calculated according to the relation
$C_2^{\rm T}/C_2^{\rm S}\simeq 7(1-n_{\rm S})$. \citet{TZ} argued,
based on this relation, that power-law models with large tilt
cannot explain the observed anisotropies. However, this 
relation is only valid for power-law inflation and the SCDM model. In
particular this is no longer true when $\Lambda \neq 0$ ($\Lambda$CDM
model). The reason is the so-called ``late integrated Sachs-Wolfe
effect'', which has been well known for a long time
\citep{KS,GSV,K}. The normalization must be performed utilizing the
power spectra themselves and not the quadrupoles in order not to
include an effect of the transfer function. In Fig.~\ref{wrongts}, we
display the $\Lambda$CDM multipole moments (for $n_{\rm S}=0.9$) in
the case where the wrong normalization is used together with the case
where the normalization is correctly calculated with the help of
Eqs.~(\ref{specsrd}) -- (\ref{specsrgw}).  The error is $\simeq 15\%$
at $\ell \simeq 200$. This weakens the mentioned argument of
\citet{TZ} and in fact questions any analysis that uses the
CMBFAST default scalar-tensor ratio together with a non-vanishing
cosmological constant.  e- Finally, CMBFAST and the
pre-July~2000 versions of CAMB calculate the low-$\ell$
multipoles in the tensorial sector inaccurately. In the case of
power-law inflation and the SCDM model they can be well
approximated by 
\begin{equation}
\label{ClTpl}
C_{\ell }^{\rm T}=
\frac{9\pi }{4}(\ell -1)\ell (\ell +1)(\ell +2)A_{\ell }(n_{\rm T}),
\end{equation}
with,
\begin{equation}
\label{defA}
A_{\ell }(n_{\rm T})\equiv \frac{400 A}{9}
\int _0^{\infty }{{\rm d}y \over y^{1 - n_{\rm T}}}
\biggl\vert \int _0^y\frac{j_2(x)j_{\ell }(y-x)}{x(y-x)^2}{\rm d}x\biggr\vert^2,
\end{equation}
where $j_{\ell }$ is a spherical Bessel function. For
$n_{\rm T}=0$, this gives in agreement with \citet{Grigw}: $C_{3}^{\rm
T}/C_{2}^{\rm T}\simeq 0.393$.  The code developed by one of us (A.R.)
reproduces this value with a precision better than $1\%$, whereas
CMBFAST gives $C_{3}^{\rm T}/C_{2}^{\rm T}\simeq 0.361$, i.e. an error
of $\simeq 8\%$. Above $\ell = 4$ both codes agree reasonably well.
This problem has been fixed in the July~2000 version of CAMB. 

\section{Test of slow-roll inflation}

We now consider the most simple and physical model for inflation
(i.e.~slow-roll inflation optimized with $\ell_{\rm opt}=200$) for the
SCDM and $\Lambda$CDM scenarios and compare its predictions with the
observational data of COBE/DMR \citep{COBE}, BOOMERanG and MAXIMA-1.
We demonstrate that a large region of the parameter space $(n_{\rm
S},n_{\rm T})$ [or equivalently $(\epsilon,\delta)$], forbidden in the
case of power-law inflation but allowed in the case of slow-roll
inflation, contains models which fit the data as good as the models
usually considered in the data analysis.  \par The data are very often
presented in terms of band-power $\delta T_{\ell}^2\equiv T_0^2\ell
(\ell +1)C_{\ell }/(2\pi )$, with $T_0 \simeq 2.7\mbox{K}$. For any
value of $\epsilon $ and $\delta$, $\delta T_{\ell}^2$ can be
approximatively expressed in terms of the band-power for $\epsilon$,
$\delta \ll 1$
\begin{equation}
\label{bandpower}
\delta T_{\ell }^2(\epsilon, \delta )\simeq 
\frac{\delta T_{\ell }^{2}(\epsilon, \delta \ll 1)}
{1+R_{10}}\biggl[1-2(2\epsilon -\delta)\ln \frac{\ell}{10}\biggr].
\end{equation}
A corresponding formula for power-law inflation has been presented in
\citet{TWL} [see remark before Eq. (33)].  For the SCDM and
$\Lambda$CDM scenarios considered here (see the introduction), we have
respectively for the first peak: $\delta T_{207}^2(\epsilon, \delta
\ll 1) \simeq 3705 (\mu\mbox{K})^2$, $\delta T_{222}^2(\epsilon,
\delta \ll 1) \simeq 6422 (\mu\mbox{K})^2$, and for the second peak
$\delta T_{497}^2(\epsilon, \delta \ll 1)\simeq 1952 (\mu\mbox{K})^2$,
$\delta T_{565}^2(\epsilon, \delta \ll 1) \simeq 3102
(\mu\mbox{K})^2$. In the previous equation, we have assumed $\delta
T_{\ell }^2(\epsilon, \delta ) \simeq \delta T_{\ell }^{{\rm
S}\,2}(\epsilon, \delta )$ which is valid only if $\ell \gg 1$ in
order for the tensorial modes to be negligible. The quantity $R_{10}$
is defined by $R_{10} \equiv \delta T_{10}^{{\rm T}\,2}/ \delta
T_{10}^{{\rm S}\,2}$ and appears because the spectrum is normalized to
the multipole $C_{10}$. At the leading order, it can be expressed as
$R_{10}\simeq - 5.31 n_{\rm T}$ and at the next-to-leading order, it
is given by
\begin{equation}
\label{r10}
R_{10}=10.62\epsilon [1+3.96\epsilon-3.85\delta -2(\epsilon -\delta)\ln(100e)].
\end{equation}
Eq.~(\ref{bandpower}) permits us to roughly understand how the
spectrum is modified when the slow-roll parameters are changed. For
fixed $2\epsilon -\delta $, i.e.~for a fixed scalar spectral index
$n_{\rm S}$, increasing $\epsilon$, i.e.~increasing the value of
$n_{\rm T}$, lowers $\delta T_{\ell}^2$. Increasing $2\epsilon
-\delta$ (i.e.  decreasing $n_{\rm S}$) while $\epsilon$ (i.e. $n_{\rm
T}$) remains constant has the same effect.  \par
\begin{figure} 
\begin{center}
\includegraphics[angle=270,width=0.7\linewidth]{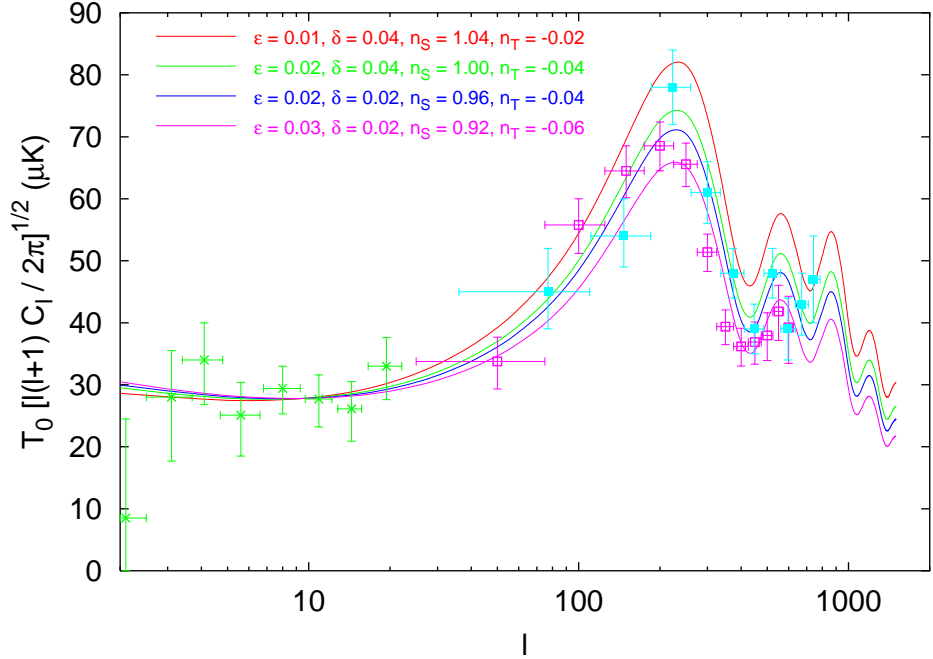}
\end{center} 
\caption{CMB band powers for slow-roll inflation in the SCDM scenario
for different values of the slow-roll parameters together with the
data points of the COBE/DMR (crosses), BOOMERanG (open boxes) and
MAXIMA-1 (filled boxes) experiments.}  
\label{scdm} 
\end{figure}

\begin{figure}
\begin{center}
\includegraphics[angle=270,width=0.7\linewidth]{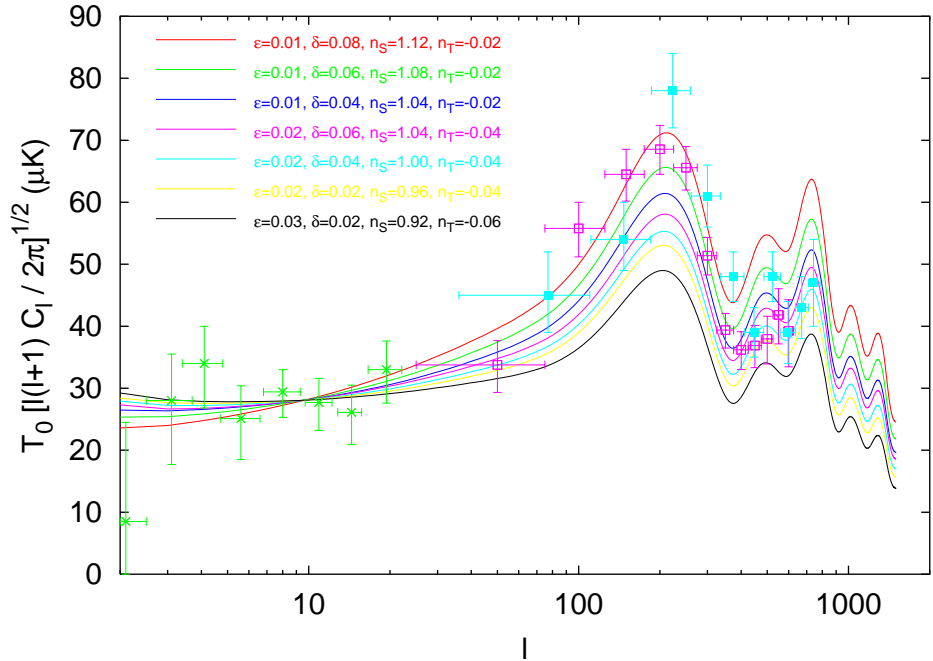}
\end{center}
\caption{As Fig.~3 but for the $\Lambda$CDM scenario.}
\label{lambdacdm}
\end{figure}

 In Figs.~\ref{scdm}
(SCDM scenario) and \ref{lambdacdm} ($\Lambda$CDM scenario), we
display the theoretical predictions of slow-roll inflation for some
values of the slow-roll parameters. Without performing a
$\chi^2$-analysis, our main conclusion is that there exist models that
reasonably fit available CMB data, which were not included in the
estimates of cosmological parameters before, in particular in the data
analysis of the recent CMB maps \citep{B98b,MAX1b}. This includes
models with $n_{\rm S}-1\neq n_{\rm T}$ and non-negligible
gravitational waves contribution. For instance, the model $\epsilon
=0.02$, $\delta =0.04$ (i.e.~$n_{\rm S}=1$ and $n_{\rm T}=-0.04$) in
the $\Lambda$CDM scenario, see Fig.~\ref{lambdacdm} goes through all
the MAXIMA-1 data points (at 1$\sigma $) but one. In this particular
case, gravitational waves represent $33\%$ of the power at $\ell =2$,
i.e.~$R_2=0.33$.  This provides a good example which violates common
(unjustified) believes about inflation. Let us stress that for both
figures we did not optimize the fits by exploring the $10\%$
resp. $4\%$ uncertainty in the calibration of the BOOMERanG and
MAXIMA-1 results, nor did we optimize the fits by varying $h$ and
$\Omega_{\rm b} h^2$ nor any other parameters.

\section{Conclusion}

It is impossible to extract the values of cosmological parameters from
the CMB anisotropy data without assumptions on the initial
spectra. For this purpose, slow-roll inflation is the best model
presently known and is consistent with presently available data.
Unfortunately, very often, only power-law inflation is considered.
The difference between both models is in general significant, which
implies that only a limited part of the space of parameters has been
correctly studied so far. Data analysis has been based on unjustified
prejudices that $n_{\rm S}$ may be greater than one in power-law
inflation, that the relation $n_{\rm S}-1=n_{\rm T}$ must hold in
general and that, gravitational waves are negligible in general. We
want to stress that a subdominant effect (as the contribution of
gravitational waves in many inflation models) is not necessarily
negligible. Although very important on the conceptual side, the
previous misconceptions were not crucial for the COBE/DMR
experiment. For the next generation of measurements, which aim to
extract cosmological parameters with a precision of a few percent,
distinguishing power-law inflation from slow-roll inflation becomes
mandatory.  We think that a correct analysis of the CMB data should
start from the spectra given in Eqs.~(\ref{specsrd})--(\ref{specsrgw})
and should be performed in the whole space of parameters
($0<\epsilon\ll 1, |\delta| \ll 1)$.  This should result in the
determination of the best $\epsilon$ and $\delta$.  The present letter
hopefully motivates more detailed tests of the most simple
inflationary scenario: slow-roll inflation.

\acknowledgments

We wish to thank A.~Lewis, U.~Seljak, and  M.~Tegmark for helpful
comments. D.~J.~S. thanks the Austrian Academy of Sciences for
financial support.

{\em Note added in web version:} 
After the proofs of our paper have been sent, version 4.0 of CMBFAST
was posted on the web. In this version the inaccuracy in the predictions 
of the tensor multipole moments has been fixed.

\clearpage

%
%
%

\end{document}